\documentclass[twocolumn,showpacs,preprintnumbers,amsmath,amssymb]{revtex4}
\usepackage{graphicx}
\usepackage{dcolumn}
\usepackage{bm}

\begin{document}
\title{ Low-momentum NN interactions and \\
all-order summation of ring diagrams
of symmetric  nuclear matter   }
\author{L.-W.\ Siu, J.\ W.\ Holt, T.\ T.\ S.\ Kuo and G.\ E.\ Brown}
\affiliation{Department of Physics and Astronomy,
Stony Brook University, NY 11794-3800, USA}
\email{thomas.kuo@stonybrook.edu}

\begin{abstract}
We study the equation of state
for symmetric nuclear matter using a ring-diagram approach in which the
particle-particle hole-hole ($pphh$) ring diagrams within a momentum
model space of decimation scale $\Lambda$ are summed to all orders.
The calculation is carried out using the renormalized
low-momentum nucleon-nucleon (NN) interaction $V_{low-k}$,
which is obtained from a bare NN potential
by integrating out the high-momentum components beyond $\Lambda$.
The bare NN potentials of CD-Bonn, Nijmegen and Idaho
have been employed.
The choice of $\Lambda$ and its influence on the
single particle spectrum
are discussed. Ring-diagram correlations at intermediate momenta
($k\simeq$ 2 fm$^{-1}$) are found to be particularly important
for nuclear saturation, suggesting the necessity
of using a sufficiently large decimation scale
so that the above momentum region is not
integrated out. Using $V_{low-k}$ with $\Lambda \sim 3$ fm$^{-1}$,
we perform a ring-diagram computation with the above potentials, which all
yield saturation energies $E/A$ and Fermi momenta $k_F^{(0)}$
considerably  larger than
the empirical values. On the other hand, similar computations with the
medium-dependent Brown-Rho scaled NN
potentials give satisfactory results
of $E/A \simeq -15$ MeV and $k_F^{(0)}\simeq 1.4$ fm$^{-1}$. The effect of
this medium dependence
is well reproduced by an empirical 3-body force of the Skyrme type.
\end{abstract}
\pacs{pacs} \maketitle

\section{Introduction}
Obtaining the energy per nucleon ($E/A$) as a function of the Fermi
momentum ($k_F$) for symmetric nuclear matter
is one of the most important problems in nuclear physics. Empirically, nuclear
matter saturates at $E/A \simeq -16$ MeV and $k_F \simeq 1.36$ fm$^{-1}$.
A great amount of effort has been put into computing the above quantities
starting from a microscopic many-body theory. For many years,
the Brueckner-Hartree-Fock (BHF) theory \cite{bethe,mach89,jwholt} was
the primary framework for nuclear matter calculations. However, BHF represents
only the first-order approximation in the general hole-line expansion
\cite{day1}. Conclusive studies \cite{day2,baldo,baldo2} have shown that the
hole-line
expansion converges at third order (or second order with a continuous
single-particle spectrum) and that such results are in good
agreement with variational calculations \cite{day3} of the binding energy
per nucleon. Nonetheless, all
such calculations have shown that it is very difficult to obtain {\it both}
the empirical saturation energy and saturation Fermi momentum simultaneously.
In fact, such calculations using various models of the nucleon-nucleon
interaction result in a series of saturation points which actually lie along a
band, often referred to as
the Coester band \cite{coest70}, which deviates significantly from
the empirical saturation point. For this reason it is now widely believed that
free-space two-nucleon interactions alone are insufficient to describe the
properties of nuclear systems close to saturation density and
that accurate results can only be achieved by introducing
higher-order effects, e.g.\ three-nucleon forces \cite{wiringa} or
relativistic effects \cite{brockmann}.

\begin{figure}[here]
\scalebox{0.5}{
\includegraphics{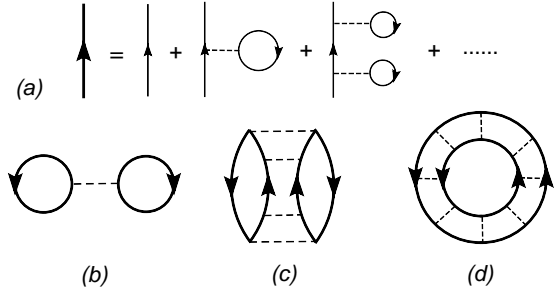}
}
\caption{\label{energy}Diagrams included in the $pphh$ ring-diagram summation
  for the ground state energy shift of symmetric nuclear matter. Included are
(a) self-energy insertions on the single-particle propagator and (b) general
$pphh$ correlations.}
\end{figure}

In the present work, we shall carry out calculations of the nuclear binding
energy for symmetric nuclear matter using a framework based on a combination
of the recently developed low-momentum NN interaction $V_{low-k}$
\cite{bogner01,bogner02,coraggio02,schwenk02,bogner03,jdholt}
and the ring-diagram method for nuclear matter of
Song {\it et al.} \cite{song87}, which is a model-space approach where the
particle-particle hole-hole ($pphh$) ring diagrams for the potential energy
of nuclear matter are summed
to all orders. In previous studies a model space of size $\Lambda$
$\sim 3$ fm$^{-1}$
was used to obtain
improved results compared with those from the BHF method. Such an improvement
can be attributed to the following
 desirable features in the ring diagram approach. First,
the ground-state energy shift $\Delta E_0$ in the BHF approach is given
by just the lowest-order
reaction matrix ($G$-matrix) diagram (corresponding to
diagram (b) of Fig.\ \ref{energy} with the dashed vertex representing $G$). It
does not include diagrams corresponding to the particle-hole excitations of the
Fermi sea. Such excitations represent the effect of long-range correlations.
In contrast, the $pphh$ ring diagrams, such as diagrams (c) and (d) in Fig.\
\ref{energy}, are included to all orders in the ring-diagram
approach. Secondly, the single-particle
(s.p.) spectrum used in the ring-diagram approach is different
from that in early BHF calculations, where one typically employed
a self-consistent s.p.\ spectrum for momenta $k \leq k_F$ and
a free-particle spectrum otherwise. Thus the s.p.\ spectrum
had a large artificial discontinuity at $k_F$.
The s.p.\ spectrum used
in the ring diagram approach is a continuous one. The importance of
using a continuous s.p. spectrum in nuclear matter theory
has been discussed and  emphasized in Ref.\cite{baldo,baldo2}.
Within the above ring diagram framework,
previous calculations \cite{jiang88} using $G$-matrix effective
interactions and $\Lambda\sim 3$ fm$^{-1}$ have yielded saturated
nuclear matter that is slightly overbound ($E/A \simeq -18$ MeV)
and that saturates at too high a density ($k_F \simeq 1.6$ fm$^{-1}$)
compared to empirical data. These results are consistent, within
theoretical errors, with calculations based on
the third-order hole-line expansion and variational methods (see Refs.\
\cite{day2,baldo,day3}). 

In the past, the above ring-diagram approach \cite{song87,jiang88} employed
the $G$-matrix interaction which is energy dependent, meaning that the
whole calculation must be done in a ``self-consistent'' way.
The calculation would be greatly simplified if this energy dependence,
and thus the self-consistency procedure, were removed.
Such an improvement has occurred in the past several
years with the development of a low-momentum NN interaction, $V_{low-k}$,
constructed from renormalization group techniques
\cite{bogner01,bogner02,coraggio02,schwenk02,bogner03,jdholt}.
As discussed in these references, the $V_{low-k}$ interaction
has a number of desirable properties, such as being nearly unique
as well as being a smooth potential suitable for perturbative
many body calculations \cite{bogner05}. 
Furthermore, $V_{low-k}$ is energy independent, making it a convenient choice
for the interaction used in ring diagram calculations of nuclear matter.

The $V_{low-k}$ interaction has been extensively used in nuclear
shell-model
calculations for nuclei with a few valence nucleons outside a closed shell.
As reviewed recently by Coraggio {\it et al.} \cite{coraggio08}, the results
obtained from such shell-model calculations are
in very good agreement with experiments.
However, applications of the $V_{low-k}$ interaction to nuclear matter
have been relatively few \cite{muether03,bogner05,jeholt07,siu2008}.
 A main purpose of the present work is to study the suitability of describing
 symmetric nuclear matter using $V_{low-k}$. A concern about such applications
is that the use of $V_{low-k}$ alone may not provide satisfactory nuclear
saturation. As illustrated in ref.\ \cite{muether03},
Hartree-Fock (HF) calculations of nuclear matter using $V_{low-k}$ with a
cutoff momentum of $\Lambda \sim 2.0$ fm$^{-1}$
 do not yield nuclear saturation--the calculated $E/A$
decreases monotonically with $k_F$ up to the decimation scale
$\Lambda$.

In this work, we carry out a ring-diagram calculation of symmetric nuclear
matter with $V_{low-k}$.
We shall show in detail that satisfactory results for
the saturation energy and saturation Fermi momentum can be obtained when one
takes into account the following two factors: a suitable choice
of the cutoff momentum and the in-medium modification of
meson masses.
As we shall discuss, ring-diagram correlations at intermediate
momenta ($k \sim 2.0$ fm$^{-1}$) have strong medium dependence and
are important for nuclear saturation. To include their
effects one needs to use a sufficiently large decimation scale
$\Lambda$ so that the above momentum range is not integrated out.
We have carried out ring-diagram calculations for symmetric nuclear
matter using $\Lambda \sim 3$ fm$^{-1}$
with several modern high-precision NN potentials, and the results
yield nuclear saturation.
However, $E/A$ and $k_F$ at saturation are both considerably
larger in magnitude than the corresponding empirical values.
Great improvement can be obtained when one takes into account medium
modifications to the exchanged mesons. Clearly mesons in a nuclear medium and
those in free space are different: the former are ``dressed'' while the latter
are ``bare''. Brown and Rho have suggested that
the dependence of meson masses on nuclear density can be described by a
simple equation known as Brown-Rho scaling \cite{brown91,brown04}:
\begin{equation}
\sqrt{\frac{g_A}{g_A^*}}\frac{m_N^*}{m_N} = \frac{m_\sigma^*}{m_\sigma} =
\frac{m_\rho^*}{m_\rho} =
\frac{m_\omega^*}{m_\omega} = \frac{f_\pi^*}{f_\pi} = \Phi(n),
\label{brs4}
\end{equation}
where $g_A$ is the axial coupling constant, $\Phi$ is a function of the
nuclear density $n$, and the star
indicates in-medium values of the given quantities. At saturation density
$\Phi(n_0) \simeq 0.8$.
 In a high-density medium such as nuclear matter,
these medium modifications of meson masses are significant and can render
 $V_{NN}$ quite different from that in free space.
 Thus, in contrast to shell model calculations for nuclei with only a few
 valence particles, for nuclear matter calculations it may be necessary to use
a $V_{NN}$ with medium modifications {\it built in}. In the present work, we
shall carry out such a ring-diagram summation using a
Brown-Rho scaled NN interaction.

The Skyrme \cite{skyrme} interaction is one of the most successful effective
nuclear potentials.
An important component of this interaction is a zero-range
3-body force, which is equivalent to a density-dependent 2-body
force.
Note that the importance of 3-body interactions in achieving
nuclear saturation with low-momentum interactions
has been extensively discussed in the literature (see ref.\ \cite{bogner05}
and references quoted therein).
In the last part of our work,
we shall study whether the density dependence from Brown-Rho scaling can be
well represented by that from an empirical density-dependent force
of the Skyrme type.

The organization of this paper is as follows. In Sections \ref{rds} and
\ref{brss} we outline our model space
$pphh$ ring-diagram calculation for the nuclear binding energy and the
concept of Brown-Rho scaling respectively. In Sec.\ \ref{resu} we present
our computational results. A brief conclusion can be found in Sec.\ \ref{conc}.

\section{Summation of $pphh$ ring diagrams}
\label{rds}

In this section we describe how to calculate the properties of
symmetric matter using the low-momentum  ring diagram  method.
We employ a momentum model space where all nucleons have momenta
$k\le \Lambda$. By integrating out the $k>\Lambda$ components,
the low-momentum interaction $V_{low-k}$ is constructed for summing
the $pphh$ ring diagrams within the model space.

The ground state energy shift $\Delta E_0 = E_0-E_0^{\rm free}$ for nuclear
 matter
 is defined as the difference between the true ground-state energy $E_0$
 and the corresponding quantity for the non-interacting system
 $E_0^{\rm free}$. In the present work, we consider $\Delta E_0$
as given by the all-order sum of the  $pphh$ ring
diagrams as shown in (b), (c) and (d) of Fig.\ \ref{energy}.

We shall calculate the all-order
sum, denoted as $\Delta E_0 ^{pp}$, of such diagrams.  Each vertex
in a ring diagram is the renormalized effective interaction
$V_{low-k}$ corresponding
to the model space $k\leq \Lambda$. It is obtained from the following
$T$-matrix equivalence method
\cite{bogner01,bogner02,coraggio02,schwenk02,bogner03,jdholt}. Let us
start with the $T$-matrix equation
\begin{multline}
  T(k',k,k^2) = V(k',k) \\
 + {\cal P} \int _0 ^{\infty} q^2 dq  \frac{V(k',q)T(q,k,k^2 )} {k^2-q^2}  ,
\label{vlk1}
\end{multline}
where $V$ is a bare NN potential. In the present work
we shall use the CD-Bonn \cite{cdbonn}, Nijmegen-I \cite{nijmegen}
 and Idaho(chiral) \cite{chiralvnn} NN potentials.
Notice that in the above equation the intermediate state momentum $q$ is
integrated from 0 to $\infty$.
We then define an effective low-momentum $T$-matrix by
\begin{multline}
T_{low-k }(p',p,p^2) = V_{low-k }(p',p) \\
 + {\cal P} \int_0 ^{\Lambda} q^2 dq \frac{V_{low-k }(p',q)T_{low-k}
 (q,p,p^2)} {p^2-q^2 },
\label{vlk2}
\end{multline}
where the intermediate state momentum is integrated from
0 to $\Lambda$, the momentum space cutoff. The low momentum interaction
$V_{low-k}$ is then obtained from the above equations by  requiring the
$T$-matrix equivalence condition to hold, namely
\begin{equation}
 T(p',p,p^2 ) = T_{low-k }(p',p, p^2 ) ;~( p',p) \leq \Lambda.
\label{vlk3}
\end{equation}
  The iteration method of Lee-Suzuki-Andreozzi
\cite{suzuki80,andre96,jdholt}  has been used in obtaining
the above $V_{low-k}$.


With $V_{low-k}$, our ring diagram calculations are relatively simple,
compared to the $G$-matrix calculations of ref.\ \cite{song87}. Within the
model space, we use the Hartree-Fock s.p.\ spectrum calculated with the
$V_{low-k}$ interaction, and outside the model space we use the free particle
spectrum. In other words,
\begin{equation}
 \epsilon_k = \left \{ \begin{array}{c}
  \hbar^2k^2/2m + \sum_{h < k_F}\langle kh|V_{low-k}|kh\rangle;
  ~~k \leq \Lambda \\[6pt]
  \hbar^2k^2/2m; ~~k>\Lambda.  \end{array} \right .
\label{sp}
\end{equation}
The above s.p.\ spectrum is medium ($k_F$) dependent.

Our next step is to solve the model space RPA equation
 \begin{multline}
\sum _{ef}[(\epsilon_i+\epsilon_j)\delta_{ij,ef}+
\lambda(\bar n_i\bar n_j -n_in_j)\langle ij|V_{low-k}|ef\rangle] \\
\times Y_n(ef,\lambda)
=\omega_nY_n(ij,\lambda);~~(i,j,e,f)\leq\Lambda, \label{rpa}
\end{multline}
where $n_a=1$ for $a \leq k_F$ and $n_a=0$ for $k>k_F$; also $\bar
n_a=(1-n_a)$.
The strength parameter $\lambda$ is introduced for calculational convenience
and varies between 0 and 1.
Note that the above equation is within the model space as indicated
by $(i,j,e,f)\leq \Lambda$. The transition amplitudes $Y$ of the above
equation can be classified into two types, one dominated by hole-hole
 and the other by particle-particle components. We use only the former,
 denoted by $Y_m$, for the calculation of the all-order sum of the
$pphh$ ring diagrams. This sum is given by \cite{song87,tzeng94,siu2008}
\begin{multline}\label{eng}
\Delta E^{pp}_0=\int_0^1 d\lambda
\sum_m \sum_{ijkl<\Lambda}Y_m(ij,\lambda) \\ \times Y_m^*(kl,\lambda) \langle
ij|V_{low-k}|kl \rangle,
\end{multline}
where the normalization condition for $Y_m$ is
$\langle Y_m|\frac{1}{Q}|Y_m\rangle=-1$ and
$Q(i,j)=(\bar n_i \bar n_j -n_in_j)$. In the above,
$\underset{m}{\Sigma}$
means
we sum over only those solutions of the RPA equation (\ref{rpa}) which are
dominated
by hole-hole components as indicated by the normalization condition.

  The all-order sum of the $pphh$ ring diagrams as indicated by diagrams
(b-d) of Fig.\ \ref{energy} is given by the above $\Delta E_0^{pp}$. Since we
use the HF
s.p.\ spectrum, each propagator of the diagrams contains the HF insertions to
all orders as indicated by part (a) of the figure. Clearly our ring diagrams
are medium dependent; their s.p.\ propagators have all-order HF insertions
which are medium dependent, as is the occupation factor
($\bar n_i \bar n_j-n_in_j$) of the RPA equation.


\section{Brown-Rho scaling and in-medium NN interactions}
\label{brss}

Nucleon-nucleon interactions are mediated by meson exchange,
and clearly the in-medium modification of meson masses is important
for NN interactions. These modifications
could arise from the partial restoration of chiral symmetry at finite
density/temperature or from traditional many-body effects.
Particularly important
are the vector mesons, for which there is now evidence from both theory
\cite{hatsuda, harada, klingl} and experiment
\cite{metag, naruki}
that the masses may decrease by approximately $10-15\%$ at normal nuclear
matter density and zero temperature. This in-medium decrease of meson masses
 is often referred to as Brown-Rho scaling
\cite{brown91,brown04}. For densities below that of nuclear matter,
it is suggested \cite{hatsuda} that the masses decrease linearly with the
density $n$:
\begin{equation}
\frac{m_V^*}{m_V} = 1- C \frac{n}{n_0},
\label{brs}
\end{equation}
where $m_V^*$ is the vector meson mass in-medium, $n_0$ is nuclear matter
saturation density and $C$ is a constant
of value $\sim 0.10-0.15$.

We study the consequences for nuclear many-body calculations by replacing
the NN interaction in free space with a
density-dependent interaction
 with medium-modified meson exchange.  A simple way
to obtain such potentials is by modifying the meson masses and relevant
parameters of the one-boson-exchange NN potentials
 (e.g.\ the Bonn and Nijmegen interactions).
The saturation of nuclear matter is an appropriate phenomenon for
studying the effects of dropping masses \cite{rapp99, jeholt07},
since the density of nuclear matter is constant and large enough
to significantly affect the nuclear interaction through the modified
meson masses.

One unambiguous prediction of Brown-Rho scaling in dense nuclear matter is the
decreasing of the tensor force component of the nuclear interaction. The two
most important contributions to the tensor
force come from $\pi$ and $\rho$-meson exchange, which act opposite to
each other:
\begin{equation}
V_\rho^T(r)=-\frac{f_\rho^2}{4\pi}m_\rho \tau_1 \cdot \tau_2 S_{12}
f_3(m_\rho r),
\end{equation}
\begin{equation}
V_\pi^T(r)=\frac{f_\pi^2}{4\pi}m_\pi \tau_1 \cdot \tau_2 S_{12}
f_3(m_\pi r),
\end{equation}
\begin{equation}
f_3(mr)=
\left(\frac{1}{(mr)^3} + \frac{1}{(mr)^2} + \frac{1}{3mr}
\right)e^{-m r}.
\end{equation}
In Brown-Rho scaling the $\rho$ meson is expected to decrease in mass at
finite density while the pion mass remains nearly unchanged due to
chiral invariance. Therefore, the overall strength
of the tensor force at finite density will be significantly smaller than that
in free space. As we shall discuss later, this decrease in the tensor
force plays an important role for nuclear saturation.



The Skyrme effective interaction has been widely used in nuclear physics
and has been very successful in describing the properties of finite nuclei
as well as nuclear matter\cite{skyrme}. This interaction has both
2-body and 3-body terms, having the form
\begin{equation}
V_{skyrme}=\sum_{i<j} V(i,j) +\sum_{i<j<k} V(i,j,k).
\end{equation}
Here $V(i,j)$ is a momentum ($\vec k$) dependent zero-range interaction,
containing two types of terms: one with no momentum dependence and the other
 depending quadratically on $\vec k$. $V(i,j)$ corresponds to a
 low-momentum expansion of an underlying NN interaction.
Its 3-body term is a zero-range interaction
\begin{equation}
V(i,j,k)=t_3\delta(\vec r_i-\vec r_j)\delta(\vec r_j-\vec r_k)
\end{equation}
which is equivalent to a density-dependent 2-body interaction
of the form
\begin{equation}
 V_{\rho}(1,2)=\frac{1}{6}t_3 \delta (\vec r_1-\vec r_2)
\rho (\vec r_{av})
\end{equation}
 with $\vec r_{av}= \frac{1}{2}(\vec r_1 +\vec r_2)$.

The general structure of $V_{skyrme}$ is rather
similar to the effective interactions based on effective field theories
 (EFT) \cite{bogner05},
with $V(i,j)$ corresponding to $V_{low-k}$ and $V(i,j,k)$
to the EFT 3-body force. The Skyrme 3-body force, however, is much simpler than
that in EFT. We shall compare in the next section the density dependent
effect generated by the medium modified NN interaction with that from
an empirical 3-body force of the Skyrme type.

\section{Results and discussions}
\label{resu}

In this section, we shall report computational results for the binding energy
of symmetric nuclear matter calculated with an all-order summation
of low-momentum $pphh$ ring diagrams. The method is already outlined and
 discussed in the above sections.
As mentioned above,
we employ a model space approach. Starting from various bare NN interactions,
we first construct the low-momentum
interactions $V_{low-k}$ with a particular choice of the cutoff momentum
$\Lambda$. The low-momentum ($<\Lambda$) $pphh$ ring-diagrams are then summed
to all-orders as given by Eq.\ (\ref{eng}) to give the binding energy.

\subsection{Single-particle spectrum and  nuclear binding energy}
\begin{figure}[here]
\scalebox{0.5}{
\includegraphics[width=13cm,angle=270]{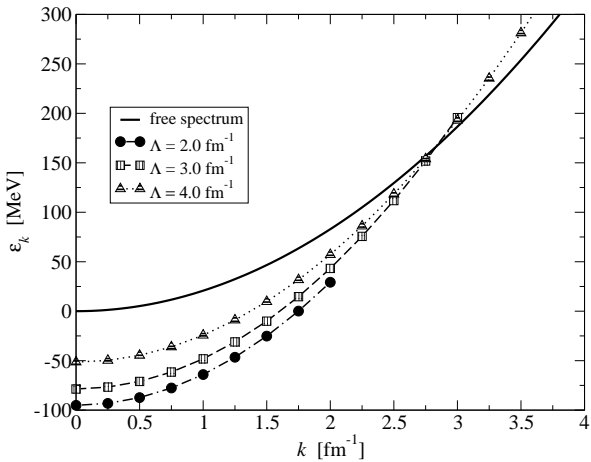}
}
\caption{\label{spen}Dependence of the model-space s.p.\ spectrum on the
  decimation scale $\Lambda$ for symmetric nuclear matter at the empirical
  saturation density. The CD-Bonn
potential is used in the construction of $V_{low-k}$.}
\end{figure}

First, we shall look carefully into the role of $\Lambda$ in
our ring-diagram calculation. Let us start with the single particle (s.p.)
energy $\epsilon_k$.  Obtaining $\epsilon_k$ is the first step
in our ring-diagram calculation. Within our model space approach, $\epsilon_k$
is given by the Hartree-Fock spectrum for $k\leq\Lambda$, while for
$k>\Lambda$, $\epsilon_k$ is taken as the free spectrum (see Eq.\ (\ref{sp})).
As emphasized before, the s.p.\
spectrum obtained in this way will in general have a discontinuity at
$\Lambda$.
Such a discontinuity is a direct consequence of having a finite model space.
 It is of much interest to study the s.p.\ spectrum as $\Lambda$ is varied.
In Fig.\ \ref{spen}, we plot the spectrum for different values of $\Lambda$
ranging from $2-4$ fm$^{-1}$.
We observed that with $\Lambda=2.0$ fm$^{-1}$, the discontinuity at $\Lambda$
is relatively large; there is a gap of about 50 MeV between the s.p.\ spectrum
just inside $\Lambda$ and that outside.  However,
this discontinuity decreases if $\Lambda$ is
increased to around 3 fm$^{-1}$. At this point,
the s.p.\ spectrum is most ``satisfactory'' in the sense of being almost
continuous.
A further increase in $\Lambda$ will result in an ``unreasonable'' situation
where the s.p.\ spectrum just inside $\Lambda$ becomes significantly higher
than that outside. This is clearly
shown in the data of $\Lambda=4.0$ fm$^{-1}$.
The above results suggest that to have a nearly continuous s.p.\ spectrum,
which is physically desirable,
it is necessary to use $\Lambda \sim 3$ fm$^{-1}$.


\begin{figure}[h!]
\scalebox{0.5}{
\includegraphics[width=13.5cm,angle=270]{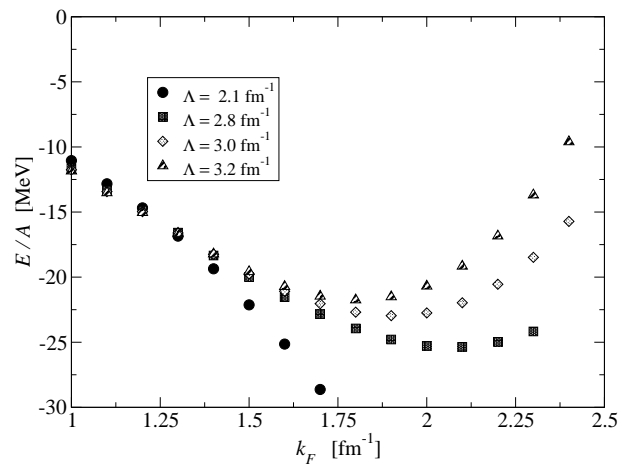}
}
\caption{\label{bea}Results for the energy per nucleon $(E/A)$ of
  symmetric nuclear matter obtained by summing up the $pphh$ ring diagrams to
  all orders. Low-momentum NN interactions, constructed from the
  CD-Bonn potential, with various cutoffs $\Lambda$ are used in the ring
  diagram summation.}
\end{figure}

Next, we shall look into the effect of $\Lambda$ on the
nuclear binding energy.
Once the s.p.\ energies are obtained, the all-order ring-diagram summation can
be carried out (see Eqs. (\ref{rpa}) and (\ref{eng})). Let us first discuss
the computational results
based on the CD-Bonn potential.
Results from various $\Lambda$ ranging from $2-3.2$ fm$^{-1}$ are
shown in Fig.\ \ref{bea}.
Let us focus on (i) the overall saturation phenomena and
(ii) the numerical values of the binding energy and the saturation momentum.

(i) We observe that the nuclear binding energy exhibits saturation
only when $\Lambda$ is $\sim 3$ fm$^{-1}$ and beyond.
This reflects the importance of ring diagrams in the intermediate
momentum region ($k\sim 2$ fm$^{-1}$). To illustrate, let us compare
the results for the cases of
$\Lambda=$ 2 and 3 fm$^{-1}$.
As indicated by Eqs.\ (\ref{vlk1}-\ref{vlk3}), $V_{low-k}$ includes
only the $k>\Lambda$
$pp$ ladder interactions between a pair of ``free'' nucleons; there is no
medium correction included. Thus the above two cases treat correlations in the
momentum region between $2$ and 3 fm$^{-1}$ differently: the former includes
for this momentum region only $pp$ ladder interactions with medium effect
neglected, while the latter
includes both $pp$ and $hh$ correlations with medium effect, such as that
from the Pauli blocking, included. Our results indicate that the medium
effect in the above momentum region is vital for saturation.

\begin{figure}[h!]
\scalebox{0.6}{
\includegraphics[width=11.5cm,angle=270]{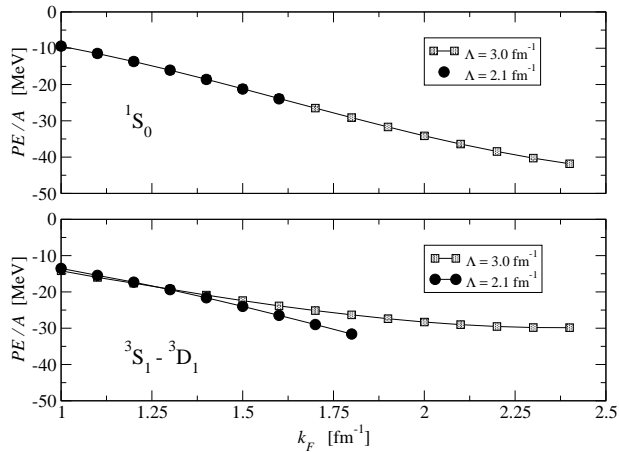}
}
\caption{\label{pea}Potential energy per nucleon $(PE/A)$ in the $^1S_0$ and
  $^3S_1-{^3D_1}$ channels of symmetric nuclear matter from summing up $pphh$
  ring diagrams to all orders. The CD-Bonn
potential is used in the construction of $V_{low-k}$.}
\end{figure}

For nuclear matter binding
energy calculations, there is no first-order contribution
from the tensor force ($V_T$); its leading contribution is second order
of the form $\langle ^3S_1|V_T\frac{Q}{e}V_T|^3S_1 \rangle$
where $Q$ stands for the Pauli blocking operator and $e$ the energy
denominator. Thus the contribution from the tensor force depends largely
on the availability of the intermediate
states; this contribution is large for low $k_F$
but is suppressed for high $k_F$. To illustrate this point,
 we plot the potential energy of nuclear matter from
the $^1S_0$ and $^3S_1 - {^3D_1}$ channels
separately in Fig.\ \ref{pea}. The behavior of the potential energy
in these two channels differ in a significant way.
The $^1S_0$ channel is practically independent of the choice of $\Lambda$,
as displayed in the upper panel of the figure. This indicates that for
this channel the effects from medium corrections and $hh$ correlations
are not important. Also the $PE/A$ from this channel does not exhibit
saturation at a reasonable $k_F$.
In the lower panel of the figure, we display
the $PE/A$ for the $^3S_1 - {^3D_1}$ channel where the tensor force is
important. As seen, $PE/A$  does not exhibit saturation
when using $\Lambda=2$ fm$^{-1}$.
On the contrary, the result using $\Lambda=3$ fm$^{-1}$
shows a clear saturation behavior. This is mainly because that
 in the former case the Pauli blocking effect is ignored
 for the momentum region $2-3$ fm$^{-1}$ while it is included for the
latter.
To have saturation, we should not integrate out
 the momentum components in the
NN interaction that are crucial for saturation.
Considering also the effect of $\Lambda$ on the s.p.\ spectrum,
we believe that $\Lambda=3.0$ fm$^{-1}$
is a suitable choice for our ring-diagram nuclear matter calculation.
Notice that a model space $\sim3$ fm$^{-1}$ has
been used in other similar ring summation calculations
using $G$-matrix effective interaction\cite{song87,jiang88}.

\begin{figure}[h!]
\scalebox{0.5}{
\includegraphics[width=11.4cm,angle=270]{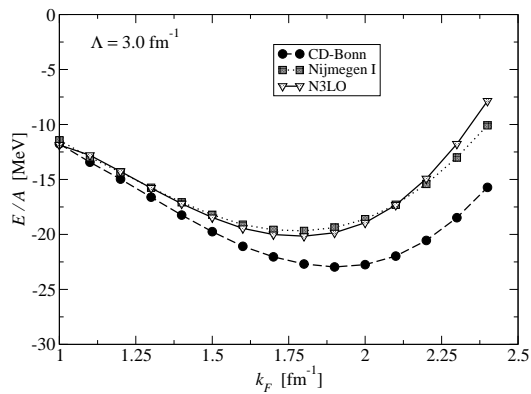}
}
\caption{\label{bea3pot}The binding energy of symmetric nuclear matter from
  the low-momentum ring-diagram summation using various NN
  potentials. A momentum-space cutoff of $\Lambda=3.0$ fm$^{-1}$ is used. }
 \end{figure}

(ii) We have performed a similar ring summation with the
 Nijmegen I and Idaho potentials. Results with $\Lambda=3.0$ fm$^{-1}$
are compared with
 that from CD-Bonn as shown in Fig.\ \ref{bea3pot}.
The saturation energies for these three potentials are located between $-19$
 and $-23$ MeV, while the saturation momentum ranges from $1.75$ to $1.85$
 fm$^{-1}$. These quantities are
 considerably larger than the empirical values of $-16$ MeV and 1.4
 fm$^{-1}$, respectively. We believe
 that improvements can be obtained if one takes into account the medium
 dependence of the NN interaction. Namely, instead of using a $V_{low-k}$
 constructed from a bare NN interaction, one should employ a $V_{low-k}$
 constructed from a ``scaled'' NN interaction according to the nuclear
 density. Below we shall report how we incorporate such effects into our ring
 diagram summation.

\subsection{Nuclear binding energy with Brown-Rho scaling}
\begin{figure}[h!]
\scalebox{0.5}{
\includegraphics[width=11.5cm,angle=270]{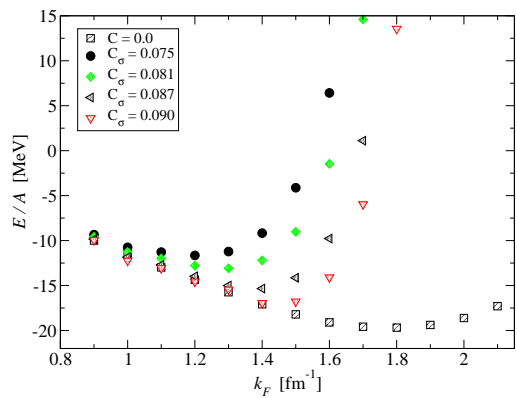}
}
\caption{\label{brbea}(color online). The binding energy of symmetric nuclear matter from
  the Brown-Rho scaled low-momentum Nijmegen II interaction using the
  ring-diagram summation with
  $\Lambda=3.0fm^{-1}$. Calculations for different choices of the $\sigma$
  meson scaling constant $C_\sigma$ are shown.}
\end{figure}

The concept of Brown-Rho scaling has already been discussed in Sec.\
\ref{brss}. The medium effects on the NN interaction resulting from the
in-medium modification of meson masses shall
have a profound effect on nuclear binding. To incorporate this in our
ring-diagram calculation we
 work with the Nijmegen potential, which is one of the pure one-boson-exchange
 NN potentials. The bare Nijmegen is first Brown-Rho
scaled (see Eq.\ (\ref{brs})) with the dropping mass ratio $C$ chosen to be
0.15. Vector meson masses in a nuclear medium have been widely studied both
theoretically and experimentally, but the $\sigma$ meson mass is not
well constrained. Previous calculations \cite{rapp99} of nuclear matter
saturation
within the Dirac-Brueckner-Hartree-Fock formalism showed that there is too
much attraction when the $\sigma$
meson is scaled according to (\ref{brs}). However, a microscopic
treatment \cite{rapp99} of
$\sigma$ meson exchange in terms of correlated $2\pi$ exchange showed that the
medium effects on the $\sigma$ are much weaker than in (\ref{brs}). Therefore,
in our ring diagram summation using the Brown-Rho scaled Nijmegen II
interaction, we employ a range of scaling parameters $C_\sigma$ between 0.075
and 0.09. Our calculations are
shown in Fig.\ \ref{brbea}. With Brown-Rho
scaling, the numerical values for
both the saturation energy and saturation momentum are greatly
improved. Whereas the unscaled potential gives a binding energy $BE/A\simeq
20$ MeV and
$k_F^0\simeq 1.8$ fm$^{-1}$, the
scaled potential gives $BE/A\simeq 14-17$ MeV and $k_F^0\simeq 1.30-1.45$
fm$^{-1}$ for a $\sigma$ meson scaling constant $C_\sigma \sim 0.08-0.09$, in
very good agreement with the empirical values. We conclude
first, that the medium dependence of nuclear interactions is crucial for a
satisfactory description of nuclear saturation and second, that within the
framework of one-boson-exchange NN interaction models one can obtain an
adequate description of nuclear matter saturation by including Brown-Rho
scaled meson masses.

\subsection{Nuclear binding energy with 3-body force of the Skyrme type}
\begin{figure}[h!]
\scalebox{0.6}{
\includegraphics[width=11.5cm,angle=270]{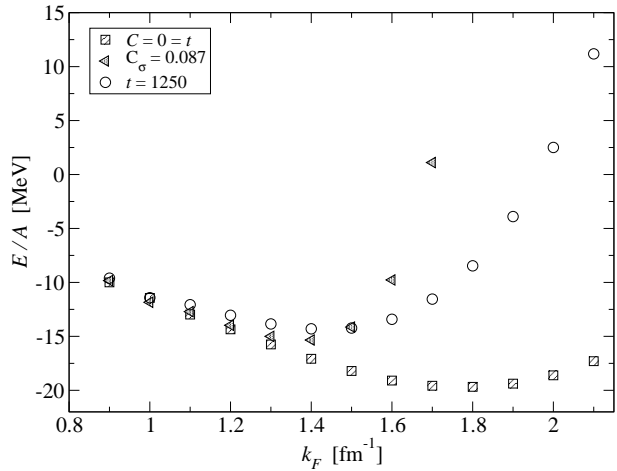}
}
\caption{\label{bea3b}The binding energy of symmetric nuclear matter from the
  low momentum ring diagram summation with $\Lambda = 3.0$
  fm$^{-1}$. Three different interactions are used: (1) the medium-independent
  Nijmegen II interaction, the Brown-Rho scaled interaction with $C_\sigma =
  0.087$, and finally the Nijmegen II interaction supplemented with a 3NF of
  the Skyrme type with $t_3=1250$.}
\end{figure}

As discussed earlier in section III, the widely used Skyrme
interaction contains a 3-body term that is equivalent
to a density-dependent 2-body interaction. It is of much
interest to study whether
our result with Brown-Rho scaled Nijmegen potential can be reproduced with the
unscaled
Nijmegen plus an effective 3-body interaction of the Skyrme type
which is characterized by a strength parameter $t_3$ (see Eq.\ (14)).
In Fig.\ \ref{bea3b} we
compare the results using $t_3=1250$ with our previous calculations using the
Brown-Rho scaled Nijmegen II potential with a $\sigma$ meson scaling constant
of $C_\sigma = 0.087$. In all calculations $\Lambda=3.0$ is used. We note that
satisfactory results for the saturation energy and Fermi momentum are obtained
using either Brown-Rho scaling or a 3NF of the Skyrme type. However, the
nuclear incompressibility is considerably larger in the case of Brown-Rho
scaling.

\section{Conclusion}
\label{conc}

We have studied the equation of state
for symmetric nuclear matter using the low-momentum nucleon-nucleon (NN)
interaction $V_{low-k}$.
Particle-particle hole-hole ($pphh$) ring diagrams within a momentum
model space $k < \Lambda$ were summed to all orders.
The significant role of the intermediate momentum range ($\sim 2.0$ fm$^{-1}$)
for nuclear saturation was discussed. We concluded that in the ring
diagram summation, having a sufficiently large model space is important
to capture the saturation effect from the intermediate momentum components.
Various bare NN potentials including CD-Bonn, Nijmegen and Idaho
have been employed, resulting in nuclear saturation with $\Lambda=3.0$
fm$^{-1}$. However, the resulting binding energy and saturation momentum are
still much larger than empirical values.
Improvement can be obtained when we take into account the
medium modification of NN interaction. We first constructed $V_{low-k}$ from
a medium-dependent Brown-Rho scaled NN
potential and then implemented this into the ring-diagram summation.
Satisfactory results
of $E/A \simeq -15$ MeV and $k_F^{(0)}\simeq 1.4$ fm$^{-1}$ could then
be obtained.
We showed that these saturation properties are
well reproduced by the first ring-diagram approach with the addition of
an empirical 3-body force of the Skyrme type.

In the future, it is of much interest to carry out a BCS calculation
on nuclear matter with $V_{low-k}$, particularly for the $^3S_1-^3D_1$ channel
where earlier calculations using bare NN interactions revealed a gap of
$~$10 MeV around normal nuclear matter densities\cite{muther05,baldo92}.
Recently, $V_{low-k}$ has been applied to obtain the equation of state
of neutron matter \cite{siu2008,schwenk03} and the $^1S_0$ pairing
gap\cite{schwenk03,hebeler07}.
A similar calculation on nuclear matter which incorporates the
tensor correlations is obviously important and we plan to investigate it in
the future.

{\bf Acknowledgement} We thank R. Machleidt for many helpful discussions.
This work is supported in part
by U.S. Department of Energy  under grant DF-FG02-88ER40388.

\vskip 1cm



\begin{thebibliography}{10}

\bibitem{bethe} H. A. Bethe, Annu. Rev. Nucl. Sci. {\bf 21} (1971) 93.
\bibitem{mach89} R. Machleidt, Adv. Nucl. Phys.  {\bf 19} (1989) 189-376.
\bibitem{jwholt} J. W. Holt and G. E. Brown, ``Hans Bethe and the Nuclear
Many-Body Problem'' in {\it Hans Bethe and His Physics}
(World Scientific, 2006, edited by G.E. Brown and C.-H. Lee).
\bibitem{day1} B. D. Day, Rev. Mod. Phys. {\bf 50} (1978) 495.
\bibitem{day2} B. D. Day, Phys. Rev. C {\bf 24} (1981) 1203.
\bibitem{baldo} H. Q. Song, M. Baldo, G. Giansiracusa, and U. Lombardo, Phys.
Lett. B {\bf 411} (1997) 237.
\bibitem{baldo2} H. Q. Song, M. Baldo, G. Giansiracusa, and U. Lombardo, Phys.
Rev. Lett. {\bf 81} (1998) 1584.
\bibitem{day3} B. D. Day and R. B. Wiringa, Phys. Rev. C {\bf 32} (1985) 1057.
\bibitem{coest70} F. Coester, S. Cohen, B.D. Day and C.M. Vincent,
Phys. Rev. {\bf C1} (1970) 769.
\bibitem{wiringa} R. B. Wiringa, V. Fiks, and A. Fabrocini, Phys. Rev. C
{\bf 38} (1988) 1010.
\bibitem{brockmann} R. Brockmann and R. Machleidt, Phys. Rev. C {\bf 42} (1990)
1965.
\bibitem{bogner01} S. K. Bogner, T. T. S. Kuo and L. Coraggio, Nucl. Phys.
{\bf A684}, (2001) 432.

\bibitem{bogner02} S. K. Bogner, T. T. S. Kuo, L. Coraggio, A. Covello
and N. Itaco, Phys. Rev. C {\bf 65} (2002) 051301R.

\bibitem{coraggio02}L. Coraggio, A. Covello, A. Gargano,
N. Itako, T. T. S. Kuo, D. R. Entem and R. Machleidt, Phys. Rev. C {\bf 66}
(2002) 021303(R).
\bibitem{schwenk02} A. Schwenk, G. E. Brown and B. Friman, Nucl. Phys.
{\bf A703} (2002) 745.
\bibitem{bogner03} S. K. Bogner, T. T. S. Kuo and A. Schwenk, Phys. Rep.
{\bf 386} (2003) 1.
\bibitem{jdholt} J. D. Holt, T. T. S. Kuo and G. E. Brown, Phys. Rev. C
{\bf 69} (2004) 034329.

\bibitem{song87} H. Q. Song, S. D. Yang and T. T. S. Kuo, Nucl. Phys.
{\bf A462} (1987) 491.
\bibitem{jiang88} M. F. Jiang, T. T. S. Kuo, and H. M\"uther, Phys. Rev. C
{\bf 38} (1988) 2408.
\bibitem{coraggio08}L. Coraggio, A. Covello, A. Gargano,
N. Itako, T. T. S. Kuo, Prog. Part. Nucl. Phys. {\bf 62} (2009) 135.
\bibitem{muether03} J. Kuckei, F. Montani, H. M\"uther and A. Sedrakian,
 Nucl. Phys. {\bf A723} (2003) 32.
\bibitem{bogner05} S. K. Bogner, A. Schwenk, R. J. Furnstahl and A. Nogga,
Nucl. Phys. {\bf A763} (2005) 59.
\bibitem{jeholt07} J. W. Holt, G. E. Brown, J. D. Holt and T. T. S. Kuo,
Nuc. Phys. A785 (2007) 322.
\bibitem{siu2008} L.-W. Siu, T. T. S. Kuo and R. Machleidt, Phys. Rev. C
{\bf 77}, 034001 (2008).
\bibitem{suzuki80} K. Suzuki and S. Y. Lee, Prog. Theor. Phys. {\bf 64} (1980)
2091.
\bibitem{andre96} F. Andreozzi, Phys. Rev. C {\bf 54} (1996) 684.
\bibitem{cdbonn} R. Machleidt, Phys. Rev. C {\bf 63} (2001) 024001.
\bibitem{nijmegen} V. G. J. Stoks, R. A. M. Klomp, C. P. F. Terheggen and
J. J. de Swart, Phys. Rev. C {\bf 49} (1994) 2950.
\bibitem{chiralvnn}D. R. Entem, R. Machleidt, Phys. Rev. C {\bf 68} (2003)
041001.
\bibitem{tzeng94}T. T. S. Kuo and Y. Tzeng, Int. Jour. Mod. Phys. E, Vol. 3,
No. 2. (1994) 523.
\bibitem{brown91}G. E. Brown, M. Rho, Phys. Rev. Lett. {\bf 66} (1991) 2720.
\bibitem{brown04}G. E. Brown, M. Rho, Phys. Rept. {\bf 396} (2004) 1.
\bibitem{rapp99}R. Rapp, R. Machleidt, J. W. Durso and G. E. Brown,
Phys. Rev. Lett. {\bf 82} (1999) 1827.
\bibitem{skyrme} P. Ring and P. Schuck, {\it The Nuclear Many-Body Problem}
(Springer-Verlag, New York, 1980), and references quoted therein.
\bibitem{hatsuda} T. Hatsuda and S. H. Lee, Phys. Rev. C {\bf 46} (1992) R34.
\bibitem{harada} M. Harada and K. Yamawaki, Phys. Rept. {\bf 381} (2003) 1.
\bibitem{klingl} F. Klingl, N. Kaiser, and W. Weise, Nucl. Phys. {\bf A624}
(1997) 527.
\bibitem{metag} D. Trnka {\it et al.}, Phys. Rev. Lett. {\bf 94} (2005) 192303.
\bibitem{naruki} M. Naruki {\it et al.} Phys. Rev. Lett. {\bf 96} (2006) 092301.
\bibitem{muther05}H. M\"uther and W. H. Dickhoff, Phys. Rev. C {\bf 72}
054313 (2005).
\bibitem{baldo92}M. Baldo, I. Bombaci, and U. Lombardo, Phys. Lett. B
{\bf 283} (1992) 8.
\bibitem{schwenk03}A. Schwenk, B. Friman and G.E. Brown, Nucl. Phys.
{\bf A713} (2003) 191.
\bibitem{hebeler07}K. Hebeler, A. Schwenk and B. Friman, Phys. Lett. B
{\bf 648} (2007) 176.








\end{thebibliography}
\end{document}